\documentclass[aps,prl,twocolumn,showpacs,superscriptaddress]{revtex4-1}
\usepackage{graphicx}
\usepackage{amsmath}
\usepackage{textcomp}
\usepackage{color}
\usepackage{xcolor}
\usepackage{upgreek}
\usepackage{tcolorbox}
\begin{document}

\title{Interaction of light carrying orbital angular momentum with a chiral dipolar scatterer}
\author{Pawe{\l} Wo{\'z}niak}
\affiliation{Max Planck Institute for the Science of Light, Erlangen, Germany}
\affiliation{Institute of Optics, Information and Photonics, Friedrich-Alexander-University Erlangen-Nuremberg, Erlangen, Germany}
\author{Israel De Le{\'o}n}
\affiliation{School of Engineering and Sciences, Tecnol{\'o}gico de Monterrey, Monterrey, Mexico}
\affiliation{University of Ottawa Centre for Extreme and Quantum Photonics, University of Ottawa, Ottawa, Canada}
\author{Katja H{\"o}flich}
\affiliation{Nanoscale Structures and Microscopic Analysis, Helmholtz-Zentrum Berlin f{\"u}r Materialien und Energie, Berlin, Germany}
\author{Gerd Leuchs}
\affiliation{Max Planck Institute for the Science of Light, Erlangen, Germany}
\affiliation{Institute of Optics, Information and Photonics, Friedrich-Alexander-University Erlangen-Nuremberg, Erlangen, Germany}
\affiliation{University of Ottawa Centre for Extreme and Quantum Photonics, University of Ottawa, Ottawa, Canada}
\author{Peter Banzer}
\email{peter.banzer@mpl.mpg.de}
\affiliation{Max Planck Institute for the Science of Light, Erlangen, Germany}
\affiliation{Institute of Optics, Information and Photonics, Friedrich-Alexander-University Erlangen-Nuremberg, Erlangen, Germany}
\affiliation{University of Ottawa Centre for Extreme and Quantum Photonics, University of Ottawa, Ottawa, Canada}
\date{\today}

\begin{abstract}
{The capability to distinguish the handedness of circularly polarized light is a well-known intrinsic property of a chiral nanostructure. It is a long-standing controversial debate, however, whether a chiral object can also sense the vorticity, or the orbital angular momentum (OAM), of a light field. Since OAM is a non-local property, it seems rather counter-intuitive that a point-like chiral object could be able to distinguish the sense of the wave-front of light carrying OAM. Here, we show that a dipolar chiral nanostructure is indeed capable of distinguishing the sign of the phase vortex of the incoming light beam. To this end, we take advantage of the conversion of the sign of OAM, carried by a linearly polarized Laguerre-Gaussian beam, into the sign of optical chirality upon tight focusing. Our study provides for a deeper insight into the discussion of chiral light-matter interactions and the respective role of OAM.}
\end{abstract}
\maketitle
%
\section{Introduction} 
\begin{figure*}[t]
\centering 
\includegraphics[width=1\textwidth]{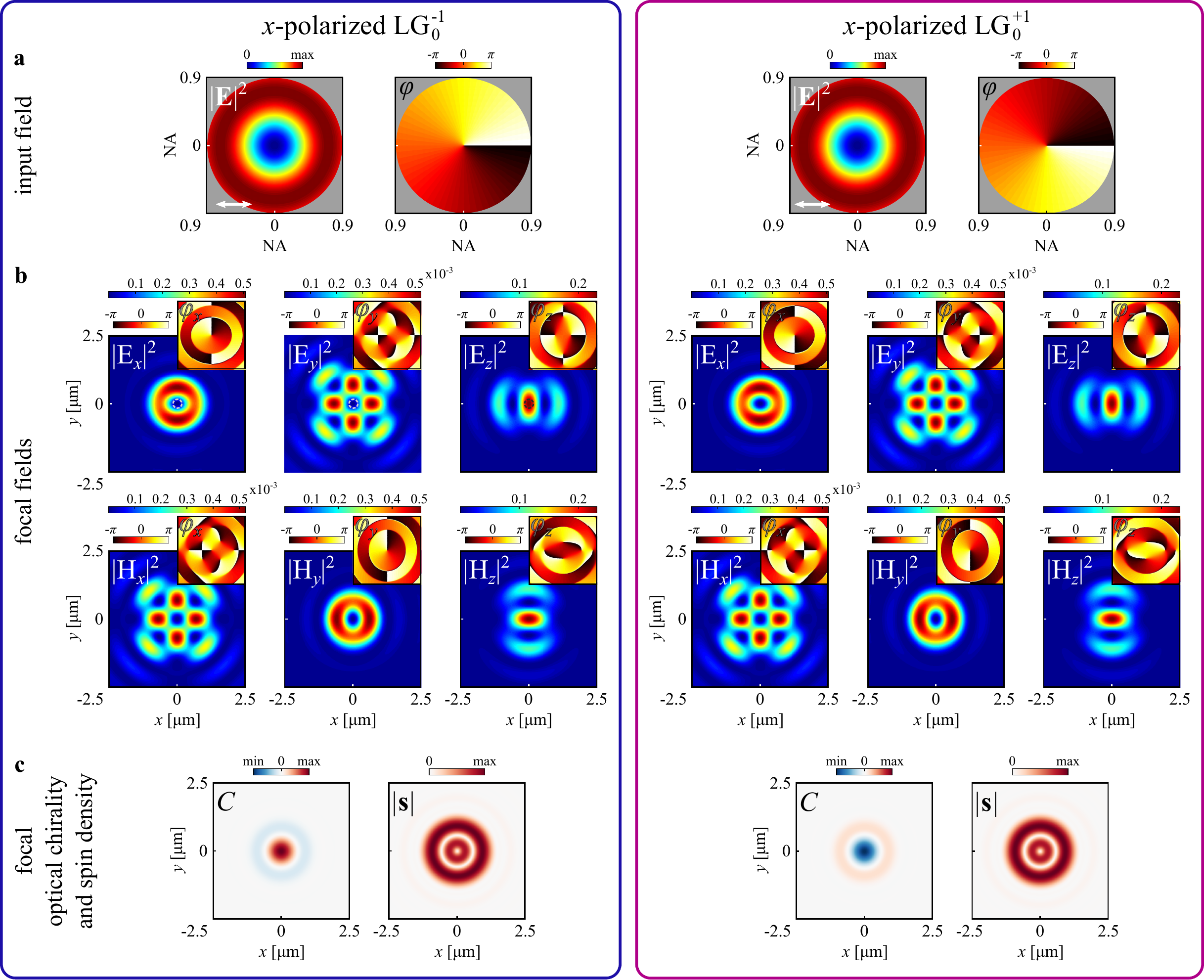} 
\caption{\textbf{$\text{LG}_{0}^{\pm1}$ beams and their field components upon tight focusing.} \textbf{a} Intensity and phase distributions of the input $\text{LG}_{0}^{\pm1}$ beams. \textbf{b} Calculated distributions of focal electric ($|\textbf{E}|^2$) and magnetic ($|\textbf{H}|^2$) field intensities (and relative phases) of tightly focused (NA = 0.9) beams. The distributions are normalized to the maximum value of the total energy density ($\frac{\epsilon_{0}}{2}|\textbf{E}|^2+\frac{\mu_{0}}{2}|\textbf{H}|^2$). The on-axis focal field comprises of longitudinal electric and magnetic fields oscillating with a phase difference $\pm\pi/2$ which corresponds to the vorticity-sense of the incoming beam. \textbf{c} Linear and dephased oscillation of $\text{E}_{z}\pm i\text{H}_{z}$ generates optical chirality $C$ of opposite sign for both input beams, but no zero spin density $\textbf{s}$ on the optical axis. The dashed circles in first line of \textbf{b} (left column) outline the top-view of the nanohelix drawn to scale.}
\label{fig:_01_focal_fields}
\end{figure*}

\vspace{-0.5cm}
Chiral molecules or artificial chiral nanostructures exhibit the inherent property of interacting differently with left-handed and right-handed circularly polarized light \cite{barron_book,gansel_2009,valev_2013,frank_2013,esposito_2015_2,wozniak_2018}. One of the possible manifestations of this spin-dependent interaction is circular dichroism (CD), i.e. the differential absorption of circular polarization states of opposite handedness, which is also utilized as an enabling feature in CD spectroscopy for enantiomeric distinction. This differential light-matter interaction is a direct consequence of the chiral geometry of the involved scatterers and a spinning light field. Artificial chiral structures have also been utilized for the conversion of incoming spin angular momentum (SAM) of light into orbital angular momentum (OAM) \cite{gorodetski_2013}. In a 3D chiral nanostructure or molecule, this phenomenon is caused by the excited chiral dipole \cite{hu_2017,wozniak_2018,eismann_2018}, consisting of parallely aligned electric and magnetic dipole moments oscillating with a phase-delay of $\pm \pi/2$, and directly resulting from the geometry of the system. The vorticity of the OAM-carrying light generated by a chiral entity illuminated with circularly polarized light depends on the relative phase between the aforementioned electric and magnetic dipoles, and hence, on the handedness of the chiral object itself \cite{wozniak_2018}. Along the same line it sounds reasonable that also the inversion of this scheme might be applicable. In other words, a chiral object illuminated with a light beam exhibiting a helical phase front (or equivalently, carrying OAM) may be able to couple to a chiral nanostructure based on the sign of the vorticity, partially converting OAM back into SAM with the helicity depending on the vorticity of the illuminating beam and the handedness of the structure. Although this concept appears consequential, it has been a common understanding that chiral media, especially when treated on the level of a pure dipolar response, are not capable of distinguishing the vorticity of the impinging light and hence, structural chirality cannot couple to OAM on the dipolar level \cite{andrews_2004,araoka_2005,loeffler_2011,giammanco_2017}. As a consequence, chiral media illuminated with linearly polarized Laguerre-Gaussian (LG) beams with an azimuthal index $l$ = 1 or $l$ = -1 would show the same transmission, reflection and absorption spectra, as opposed to the differential absorption (i.e., CD) observed for illumination with left- and right-handed circularly polarized light. A simple argument for this conclusion is the fact that OAM originates from the spatial phase distribution of a light beam or field. Thus, OAM could be labeled as a non-local property of a light beam, in contrast to SAM, which is a local feature related to the polarization. This poses an apparent contradiction with the simple idea of inverting the spin-to-orbit conversion mediated by a chiral nanostructure as described above.\\
Here, we show that an individual chiral dipolar nanostructure is in fact capable of sensing the vorticity of the impinging light beam. It can distinguish between light beams carrying phase vortices of opposite charges similar to its capability of sensing the handedness or sign of the SAM of a circularly polarized light beam (see, e.g., \cite{wozniak_2018} and references therein).\\
In the scheme discussed here, we take advantage of non-paraxial propagation to create non-zero optical chirality or helicity density \cite{tang_2010} from a linearly polarized LG beam of charge $l$ = 1 or $l$ = -1, which possesses no SAM before being focused (see Fig.~\ref{fig:_01_focal_fields}). Thus, a chiral nanostructure placed in the focal field indirectly interacts with the field's OAM through the relative phase between the electric and magnetic longitudinal field components. Specifically, we show that the sign of the phase-charge of an OAM-carrying light beam gets encoded in the phase delay between the longitudinal electric and magnetic field components on the optical axis in the focal plane and, equivalently, in the sign of the optical chirality of the field \cite{tang_2010} formed there. In turn, the dipole-like chiral nanostructure can interact differently with this optical chirality depending on its own geometrical handedness and the chiral dipole it supports at the fundamental resonance. In contrast to the well-known spin-to-orbit coupling in tightly focused circularly polarized light beams \cite{zhao_2007,bliokh2015}, we convert OAM to local non-zero optical chirality.
\section{Interaction of a dipolar chiral scatterer with tightly focused $\text{LG}_{0}^{\pm1}$ beams}
\label{sec:_theory}
\begin{figure*}[t]
\centering 
\includegraphics[width=\textwidth]{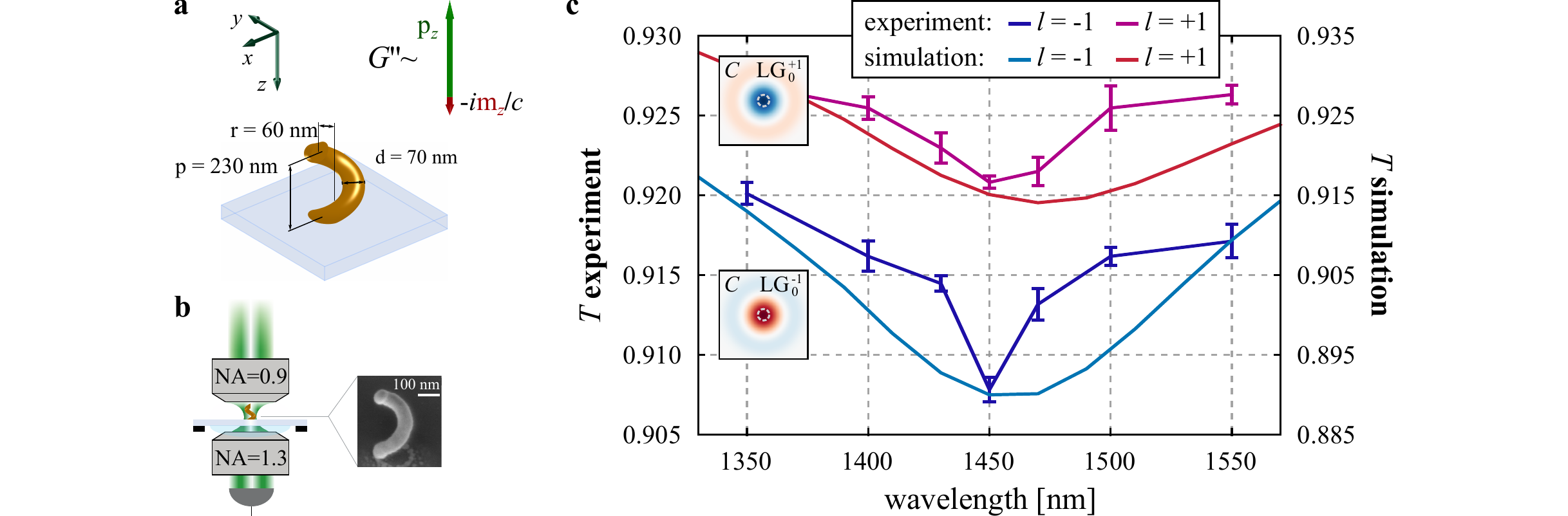}  
\caption{\textbf{Measurement scheme and the results.} \textbf{a} Simplified sketch of the experimental setup \cite{banzer_2010,wozniak_2018} for probing the vorticity of the incoming beam with a chiral scatterer and scanning-electron micrograph of the utilized nanostructure. \textbf{b} Dimensions of the nanohelix. The strength of its chiroptical response can be described by the chiral dipole oscillating along the $z$-direction (helix axis). \textbf{c} Experimental and simulation spectra of the fundamental resonance of the nanohelix as a function of the vorticity of the incoming beam. The insets depict the focal optical chirality and the relative size and position of the nanohelix in the focal plane.}
\label{fig_02_spectra}
\end{figure*}

We start by analyzing the above-mentioned phenomenon of OAM generation by a chiral object. In the interaction of circularly polarized light with a sub-wavelength-sized chiral structure (see Fig.~\ref{fig_02_spectra}a), a chiral dipole is excited as a consequence of the chiral geometry of the illuminated structure. A chiral dipole is formed by parallely aligned electric and magnetic dipole moments oscillating with a phase delay \cite{hu_2017,wozniak_2018,eismann_2018}. The emission of a chiral dipole is, therefore, ruled by the radiation of an out-of-phase superposition of parallel electric and magnetic dipole moments. The emission of an electric and a magnetic dipole moment observed around the dipole axis is radially and azimuthally polarized light, respectively. Hence, the aforementioned superposition is equivalent to a circularly polarized LG-beam of charge $|l|$ = 1 with the sign of the spin and OAM depending on the relative phase between the dipoles and the handedness of the chiral dipole. We consider tight focusing of linearly $x$-polarized LG beams of zero radial order and azimuthal order $l$ = $\pm$1 to exclude any possible influence of SAM from the input beam on the interaction between the field and the chiral nanostructure. We therefore start with a paraxial beam (propagating along the $z$-axis) carrying OAM only. The electric field distribution in the back focal plane of the focusing lens can be expressed as (see Fig.~\ref{fig:_01_focal_fields}): 
\begin{equation}
\textbf{E} = \text{E}_{0}\frac{r}{w_{0}}e^{-\frac{r^{2}}{w_{0}^{2}}}e^{\pm i l\phi}\hat{\textbf{e}}_{x}\text{,}
\label{eq:_lg_beam}
\end{equation}
where $r$ and $\phi$ are the radial and azimuthal coordinates, respectively, and $w_{0}$ is the equivalent beam waist. The focal fields can be calculated using vectorial diffraction theory \cite{richards_1959,novotny_2006}. Figure~\ref{fig:_01_focal_fields} shows the focal fields calculated for the parameters used in the experiment: $w_{0}$ = 2 mm, $NA$ = 0.9 $f$ = 2 mm and $\lambda$ = 1450 nm, with $f$ and $\lambda$ the focal length of the focusing system and the wavelength of the beam, respectively. In the following, we are mainly interested in the structure of the focal field on and near the optical axis in the focal plane. On the optical axis, the electromagnetic field is purely longitudinal (transverse components cross zero), with these axial electric and magnetic field components being dephased by exactly $\pm \pi/2$. A sub-wavelength-sized nanostructure placed at this position in the focal plane will be excited by these field components. This configuration resembles the structure of a chiral dipole as discussed above. We therefore expect efficient coupling to a chiral nanostructure, if the relative phase between the electric and magnetic field components matches the relative phase of the electric and magnetic dipole moments forming the chiral dipole mode. The latter is dictated by the geometrical handedness of the chiral nanostructure.\\
The aforementioned focal fields give rise to non-zero optical chirality, defined as (see Fig. \ref{fig:_01_focal_fields}c) \cite{lipkin_1964,tang_2010}: 
\begin{equation}
C=-\frac{\omega}{2c^{2}}\Im\left[\textbf{E}^{*}\cdot\textbf{H}\right]\text{,}
\label{eq:_optical_chirality}
\end{equation}
providing for a quantitative measure of how strongly chiral an electromagnetic field is. Importantly, the sign of the vorticity of the input beam is manifested by the phase relation between the longitudinal electric and magnetic fields as shown in Fig.~\ref{fig:_01_focal_fields}b. Thus, for incoming beams of opposite charge $l$ = $\pm$1, $C^{+1}$ = $-C^{-1}$ $\neq$ 0. In addition, the SAM (density) defining the local degree of circular polarization is zero on the optical axis, owing to the chosen symmetry (see Fig.~\ref{fig:_01_focal_fields}). It should be stressed here that the input beam (\eqref{eq:_lg_beam}) possesses neither a non-zero optical chirality nor SAM at any point in the back focal plane. Hence, the observed non-zero optical chirality on the optical axis in the focus can be solely attributed to the formation of electric and magnetic longitudinal fields with their relative phase ruled by the sign of the OAM.\\
In previous studies \cite{andrews_2004,araoka_2005,loeffler_2011,giammanco_2017} it was shown that a dipolar chiral object (e.g. a molecule) cannot show differential interaction with linearly polarized LG beams of opposite azimuthal index. More recently, it was proposed that the OAM of paraxially propagating LG modes can engage in chiral light-matter interactions via quadrupolar responses \cite{quinteiro_2017,forbes_2018,andrews_2018,kerber_2018}. Also, the role of longitudinal fields in chiral interactions was emphasized recently \cite{rosales_2012,quinteiro_2017}.\\
With our scheme we now show experimentally and theoretically that tightly focused linearly polarized $\text{LG}_{0}^{\pm1}$ beams provide the necessary condition for a dipole-like chiral particle to scatter differently for different vorticities of the input beam via the creation of longitudinal field components on the optical axis and the resulting optical chirality. Hence, we take advantage of non-paraxial propagation of structured light and the corresponding longitudinal field components created at the focal plane \cite{rosales_2012,eismann_2018}.\\
In our experiment, we utilize a plasmonic nanohelix as a prototypical chiral scatterer (see Fig.~\ref{fig_02_spectra}a). In our previous study \cite{wozniak_2018} it has been shown that the fundamental resonance of this single-loop (right-handed) plasmonic nanohelix can be approximated by the dominating longitudinal components of the electric and magnetic dipoles $\text{p}_{z}-i\text{m}_{z}$. In this context, $\text{p}_{z}$ and $\text{m}_{z}$ refer to the real and positive-valued amplitudes of the $z$-components of the electric and magnetic dipoles, respectively (see Fig.~\ref{fig_02_spectra}a). It can be therefore treated as a dipole-like chiral nanostructure. Since the nanohelix is optically chiral, it features a non-zero cross-polarizability $G''\propto -\Im\left[\textbf{p}^{*}\cdot\textbf{m}\right] \neq$ 0 \cite{barron_book,hu_2017}. We note that the phase delay of $-i$ and thereby also the handedness of the chiral dipole at the fundamental resonance is defined by the right-handed geometry of the structure studied here. Therefore, only one of the tightly focused $\text{LG}_{0}^{\pm1}$ beams will be able to interact with the helix by exciting the corresponding chiral dipole.\\
\section{Experimental results}
\label{sec:_expermient}
To experimentally (and numerically) verify the concept outlined above, we use a gold nanohelix as a sub-wavelength-sized chiral dipolar scatterer (see Fig.~\ref{fig_02_spectra}a-b). It was fabricated by electron-beam-induced deposition \cite{hoeflich_2011} combined with a subsequent grazing incidence metal coating process, resulting in a core-shell gold nanohelix. Using plane-wave-like circularly polarized excitation, the fundamental optical properties of this nanostructure were studied in detail in an article published recently \cite{wozniak_2018}. The resonance wavelength of the fundamental resonance is 1450 nm. To prove differential interaction of the nanohelix with the incoming light field featuring opposite vorticities, we measured the transmittance spectra around the fundamental resonance using a custom-built optical setup (see simplified sketch in Fig.~\ref{fig_02_spectra}b, and more details in Refs.~\cite{banzer_2010,wozniak_2018}). The helix was excited with tightly focused linearly $x$-polarized $\text{LG}_{0}^{\pm1}$ beams ($NA$ = 0.9 and $w_{0}/f$ = 1). The linearly polarized LG modes were generated using right- and left-handed circularly polarized Gaussian beams transmitted through a $q$-plate of charge -1/2 \cite{bomzon_2001,marrucci_2006} and a linear polarizer. Subsequently, the spatial modes were cleaned with a Fourier filter and focused onto the sample. The nanohelix fabricated on a glass substrate was placed on a 3D piezo-stage, which allowed for precise positioning of the structure on the optical axis in the focal plane. The transmitted and forward-scattered light was collected by a second high-NA (1.3) objective and detected using a photo-diode. We note that for a proof-of-principle experimental demonstration we measured only the total transmitted power, which should differ for different signs of OAM of the input beams.\\ 
Figure~\ref{fig_02_spectra}c shows the transmittance ($T$) spectra of the nanohelix measured wavelength-by-wavelength in the vicinity of its fundamental resonance. As explained, the strength of the interaction depends critically on the vorticity of the incoming beam. The recorded spectra show that the tightly focused $\text{LG}_{0}^{-1}$ beam can couple more efficiently to the chiral dipole mode of the nanohelix at the fundamental resonance. This is a direct consequence of the longitudinal field components and their relative phase ($\text{E}_{z}- i\text{H}_{z}$) matching the chiral dipole handedness ($\text{p}_{z}-i\text{m}_{z}$; see Fig.~\ref{fig_02_spectra}b) of the right-handed structure investigated here. The corresponding simulations (based on finite-difference time-domain method) resemble the experimental results very well (see solid continuous lines in Figure~\ref{fig_02_spectra}c). The presented spectra provide experimental and numerical evidence that a dipolar chiral scatterer can distinguish the vorticity of an incoming beam carrying OAM via the generation of non-zero optical chirality by tight-focusing.\\
Additional simulations (not shown here) also verify that by changing the handedness of the nanohelix, also the transmission properties change accordingly and stronger coupling is observed for an $\text{LG}_{0}^{1}$ beam. To also make sure that the observed differential transmission is indeed a direct consequence of the sign of the on-axis optical chirality, and, hence, the OAM of the input beams, we also checked the influence of the rotation angle of the helix about the optical axis. We find that the relative orientation of the incoming linear polarization state (and corresponding off-axis transverse field components in the focus) with respect to the nanohelix of finite length twisting around the $z$-axis does not play an important role. Rotating the nanohelix about the $z$-axis only changes slightly the strength of the observed differential transmission (see Supplementary Material).\\
It is interesting to note here that also the first higher-order resonance of the nanohelix (at $\lambda$ = 840 nm) can be described as a system of coupled electric and magnetic dipoles with $G''\neq0$ \cite{wozniak_2018}. In this case, however, the chiroptical response of the helix, in first approximation, is dominated by transverse dipole moments $\text{p}_{x}+i\text{m}_{x}$ aligned orthogonal to the optical axis of the system. For the excitation scheme based on the on-axis longitudinal fields presented above, no efficient coupling at the wavelength of the first higher-order resonance can be achieved (see Supplementary Material for details). This is contrary to the response of the helix to plane-wave-like circularly polarized excitation, which has a better overlap with the $x$-polarized chiral dipole at the first higher-order resonance.
\section{Conclusion}
\label{sec:_conclusion}
In summary, we have studied the role of OAM in chiral light-matter interactions on the level of an individual dipolar chiral nanostructure. The tight focusing of linearly polarized $\text{LG}_{0}^{\pm1}$ modes carrying no SAM but only OAM (of charge $|l|$ = 1) results in the creation of non-zero optical chirality peaking on the optical axis in the focal plane. The twisting sense of phase fronts, and, equivalently, the sign of the OAM of the incoming light field define the sign of the focal optical chirality. The corresponding field couples preferentially to the nanohelix's fundamental chiral mode if the sign of the optical chirality in the focal field (dictated by the vorticity of the input field) matches that of the relative phase between the electric and magnetic components of the nanohelix's chiral mode (dictated by the helix's handedness). In contrast to recent studies \cite{forbes_2018,andrews_2018}, the vorticity of the incoming beam was sensed via dipolar interactions not involving any higher order multipoles. To prove our concept experimentally, we investigated the differential transmission of $\text{LG}_{0}^{\pm1}$ beams tightly focused onto a single plasmonic nanohelix. Due to the sign relation between the incoming OAM and the focal optical chirality, the nanostructure was able to unambiguously recognize the vorticity of the incoming light field. Our study constitutes a new route for tailoring the chiral light-matter interactions at the nanoscale. In addition, this study sheds new light on the discussion of the role of OAM in chiral light-matter interactions. The polarization rearrangement of the focal fields caused by non-paraxial propagation is a straightforward way of engineering focal fields and optical chirality at the nanoscale, and for studying selectively the individual chiral modes of an arbitrary structure.
\section*{Funding Information}
Helmholtz Association, Helmholtz Postdoctoral Fellowship (PD140); CONACyT -- DAAD (Proalmex) grant under the project No. 267735; Project PPP Mexiko 2j16 (project-ID: 57274178) supported by DAAD with funds provided by the Federal Ministry of Education and Research (BMBF); Max Planck--University of Ottawa Centre for Extreme and Quantum Photonics.
\section*{Acknowledgments}
The authors thank Sergey Nechayev and Martin Neugebauer for fruitful discussion.
\bibliography{biblioteca}
\clearpage
\onecolumngrid

\begin{tcolorbox}
\centerline{\textbf{{Interaction of light carrying orbital angular momentum with a chiral dipolar scatterer}}}
\vspace{2 mm}
\centerline{\small{SUPPLEMENTARY INFORMATION}}
\end{tcolorbox}
\renewcommand{\thefigure}{S\arabic{figure}}
\setcounter{figure}{0}
\vspace{-0.3cm}
\section{Interaction of the nanohelix with tightly focused \textit{x}- and \textit{y}-polarized $\text{LG}_{0}^{\pm1}$ beams}
Tight focusing of $\text{LG}_{0}^{\pm1}$ beams generates focal fields, which can be described on the optical axis solely by the longitudinal field components $\text{E}_{z}\pm i\text{H}_{z}$. Transverse fields components ($x$ and $y$) are strictly zero on-axis. Hence, the on-axis field and optical chirality \cite{tang_2010} stem from the depolarization effect (due to tight focusing). A point-like scatterer would, hence, interact only with the longitudinal electromagnetic field when placed on the optical axis. The nanohelix placed at this position, however, with its outer-most radius of 95 nm, also overlaps partially with the transverse components rising off-axis. We note that at the edges of the nanohelix, the amplitude of the focal $x$-polarized electric field component is comparable in strength with respect to the $z$-polarized field peaking on the optical axis. Figure S1a shows the differential transmission ($T$) of tightly focused $x$- and $y$-polarized ($NA$ = 0.9) $\text{LG}_{0}^{\pm1}$ beams, respectively. Indeed, the chiroptical response of the nanohelix is stronger for $x$-polarized beams than for $y$-polarized beams for a fixed orientation of the helix. Yet, the small difference for the two orthogonal linear polarization states, shown in Fig. \ref{fig_01_si}a, indicates that the structure is predominantly excited by the longitudinal fields $\text{E}_{z}\pm i\text{H}_{z}$, as the chiral dipole of the fundamental resonance is dominated by its longitudinal component (see Fig. 2a in the main text and Figs 3-4 in Ref. \cite{wozniak_2018}).
\vspace{-0.3cm}
\section{Selective excitation of the chiral dipoles}
\begin{figure*}[h] 
\centering 
\includegraphics[width=\textwidth]{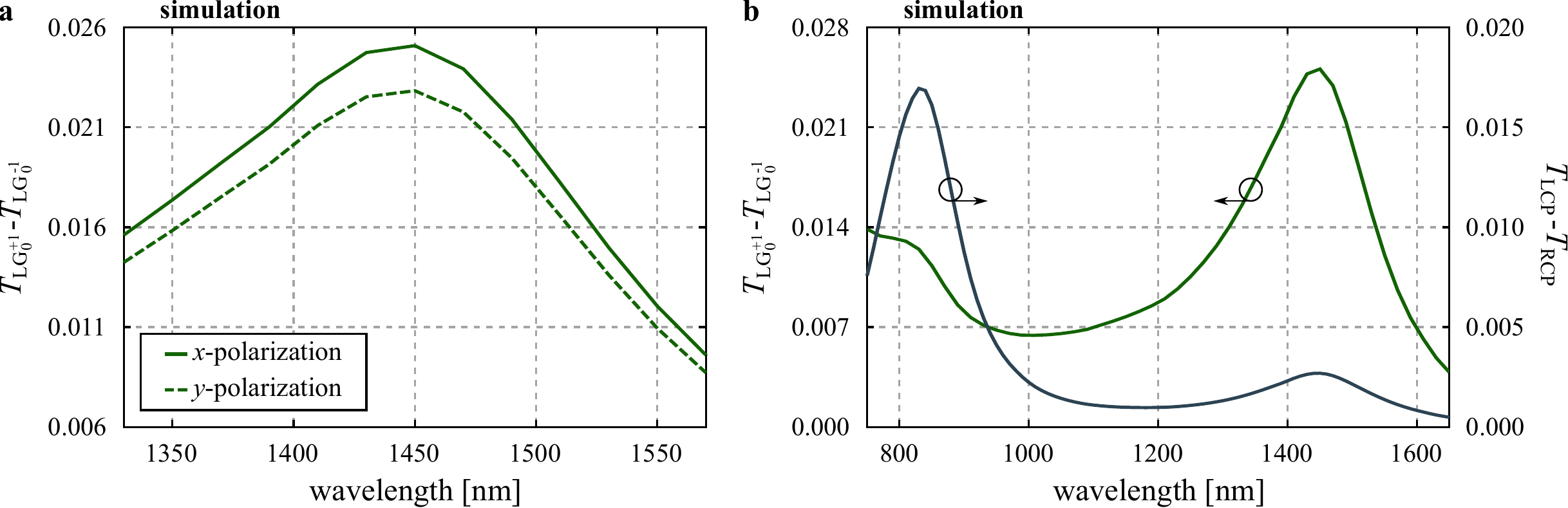}  
\caption{\textbf{Differential transmission of an individual nanohelix for different illumination schemes.} \textbf{a} Differential transmission of tightly focused \textit{x}- and \textit{y}-polarized $\text{LG}_{0}^{\pm1}$ beams around the fundamental resonance of the nanohelix ($\lambda$ = 1450), for the structure placed on-axis and oriented as depicted in Fig. 2a in the main text. \textbf{b} Differential transmission of tightly focused \textit{x}- $\text{LG}_{0}^{\pm1}$ beams and  weakly focused left-handed (LCP) and right-handed (RCP) circularly polarized Gaussian beams in the spectral range covering the fundamental resonance and the first higher-order resonance ($\lambda$ = 840 nm).}
\label{fig_01_si}
\end{figure*}

In Ref. \cite{wozniak_2018}, it was shown that the fundamental ($\lambda$ = 1450 nm) and first higher-order ($\lambda$ = 840 nm) resonances of the investigated single-loop helix are dominated by chiral dipoles. Since the two dipoles can be described, in first approximation, as $\text{p}_{z}-i\text{m}_{z}$ and $\text{p}_{x}+i\text{m}_{x}$, respectively, a stronger coupling to the fundamental resonance is expected with tightly focused $\text{LG}_{0}^{\pm1}$ beams and an on-axis placement of the nanohelix. Corresponding FDTD calculations, extended over the wavelength range from 750 nm to 1650 nm, indeed confirm a weaker differential response of the helix excited with $\text{E}_{z}\pm i\text{H}_{z}$ at the first higher-order resonance ($\lambda$ = 840 nm) compared to the fundamental resonance at $\lambda$ = 1450 nm (see green solid line in Fig. \ref{fig_01_si}b). On the contrary, a circularly polarized plane-wave-like illumination at normal incidence results in a stronger differential signal caused by the dominating transverse chiral dipole components of the first higher-order mode (see gray solid line in Fig. \ref{fig_01_si}b and Ref. \cite{wozniak_2018}). \\ 

\end{document}